\documentclass[twocolumn]{aastex62}
\usepackage{indentfirst}
\begin{document}
\title{\Large \textbf{Constraining cosmological parameters in FLRW metric with lensed GW+EM signals}}
 %% \footnote{Released on August, 8th, 2018}
%%\correspondingauthor{Fan}
\correspondingauthor{X. Fan}
\email{fanxilong@outlook.com}

\author{YUFENG LI}
\affil{Key Laboratory for Computational Astrophysics, 
National Astronomical Observatories, Chinese Academy of Sciences, 
Beijing 100012, China}
\affil{School of Astronomy and Space Science, University of Chinese Academy of Sciences, Beijing 100049, China}

\author{XILONG FAN}
\affil{School of Physics and Technology, Wuhan University, Wuhan 430072, China}
\affiliation{Department of Physics and Mechanical and Electrical Engineering, \\
Hubei University of Education, Wuhan 430205, China. }

\author{LIJUN GOU}
\affil{Key Laboratory for Computational Astrophysics, 
National Astronomical Observatories, Chinese Academy of Sciences, 
Beijing 100012, China}
\affil{School of Astronomy and Space Science, University of Chinese Academy of Sciences, Beijing 100049, China}

\nocollaboration

\begin{abstract}
We proposed a model-independent method to constrain the cosmological parameters  using the Distance Sum Rule of the FLRW metric  by  combining the time delay distances  and  the comoving distances through a multi-messenger approach. The time delay distances are measured from lensed gravitational wave~(GW) signals together with their corresponding electromagnetic wave~(EM) counterpart, while the comoving distances are obtained from a parametrized fitting approach with independent supernova observations.  With a series of simulations based on Einstein Telescope,  Large Synoptic Survey Telescope  and The Dark Energy Survey,  we find that only 10 lensed GW+EM systems can achieve the constraining power comparable to and even stronger than 300 lensed quasar systems due to more precise time delay from lensed GW signals. Specifically, the cosmological parameters can be constrained to ~$k=0.01_{-0.05}^{+0.05}$ and ~$H_0=69.7_{-0.35}^{+0.35}$ (1$\sigma$)\footnote{The uncertainty is usually given in 1$\sigma$ unless otherwise stated.}. Our results show that more precise time delay measurements could provide more stringent cosmological parameter values, and lensed GW+EM systems therefore can be applied as a powerful tool in the future precision cosmology. 
\end{abstract}

\section{Introduction} \label{sec:intro}
In 2015, Laser Interferometer Gravitational Wave Observatory (LIGO) announced the first gravitational wave event GW150914, indicating a new window is opened for astronomical observations. Later in 2017, both the gravitational wave (GW) and its electromagnetic (EM) counterpart of the binary neutron star merger GW170817 marked the arrival of multi-messenger astronomy \citep{2016PRL...116...061102L}. Combining the advantages of different messengers, multi-messenger astronomy will provide us more information of the Universe and thus enhance our understanding of the physical nature of the Universe.

One of the applications is to test the cosmological metric and to constrain the cosmological parameters. In cosmology, the Friedmann-Lema\^{\i}tre-Robertson-Walker (FLRW) metric is the most basic hypothesis, which describes a homogeneous, isotropic universe. In this respect, testing its validity is of great significance. There have been a number of methods suggested to test the FLRW metric. \citet{2008PRL...101...011301L} proposed to test it by comparing the observational measurements of the expansion rate and cosmological distances. \citet{2014JCAP...03...035L} suggested to use the  parallax distance and angular-diameter distance. And at a later time, \citet{2015PhRvL...115...101301L} found that lensing systems with the independent supernova observations can be used to test the FLRW metric using the equivalent Distance Sum Rule~(DSR). If the DSR is violated, the FLRW metric can be ruled out. However, if the data is consistent with the FLRW metric, we can go a step further to constrain the cosmological parameters~\citep{2015PhRvL...115...101301L}. In this work, we assume the FLRW metric is valid and no  other  cosmology model is  adopted.

Besides, a number of studies on constraining the cosmic curvature by CMB and BAO have been flourishing over the last decades~\citep{2005ApJ...633...560L, 2006PhRvD...74...123507, 2016AA...594...A13}. Later for the same research purpose, \citet{2010PhRvD...81...083537L} used the Hubble rate measurements together with supernovae distances; \citet{Mortsell2011} combined the distance and look-back time observations; \citet{2014PhRvD...90...023012L} proposed data binning with direct error propagation, the principal component analysis, the genetic algorithms and the Pad\'e approximants; \citet{2016PhRvD...93...043517L} applied a model-independent smoothing technique, Gaussian processes. More recently, \citet{2017ApJ...839...70L} developed a new model-independent approach to test the FLRW metric and to constrain cosmic curvature using both the strong-lens time-delay systems and independent supernovae observations, based on the Distance Sum Rule, and it turns out to be much more accurate than the previous works.

Detection of gravitational waves accompanied by electromagnetic counterparts gives us some new inspirations. For example, the strong lensed GW-EM systems have been proposed as powerful astronomy tools in a series of studies~\citep[e.g.][]{2017PhRvL.118i1102F,2017NatCo...8.1148L,2017PhRvD..95f3512B,2017PRL...118...091101,2014JCAP...10...080L,2017MNRAS...465...4634L,2013JCAP...10...022L,2018MNRAS...472...2906L}. In \citet{2015JCAP...12...006L}, they concluded that, once the third-generation GW detectors (e.g., the Einstein Telescope, ET) start to operate, $10^4$-$10^5$~GW events will be detected per year according to their sensitivity, and 50-100 out of them are expected to be lensed. The time-delay measurements from lensed GW can be quite accurate with ignorable observation error, and the time-delay is an important information for cosmology research with lensing systems. Thus lensed GW-EM system could provide stringent constraints on cosmological parameters~\citep{2017NatCo...8.1148L}. Therefore it's reasonable to consider that combining both the redshift and the Fermat potential difference observed from lensed EM with the high-accuracy time delay obtained from lensed GW, we can put tighter constraints on the cosmological parameters under the Distance Sum Rule of the FLRW metric. To achieve this purpose, we carried out a  series of simulations, and present the results in this paper.

This paper is organised as follows. In Section 2, we will describe the methodology. The simulation and fitting results will be shown in Section 3. And a brief summary to the results is presented in Section 4. %\textbf{The speed of light c = 1.}

\section{methodology} \label{sec:floats}
The method is based on the time delay distance $D_{\Delta t}$, which is usually defined as:
\begin{equation}\label{dt}
D_{\Delta t} = \frac{D_A(z_l)D_A(z_s)}{D_A(z_l,z_s)},
\end{equation}
where $z_l,z_s$ are redshifts at lens and source, respectively. 
By equating the observational and theoretical time delay distance, we can obtain the cosmic curvature k given the observed quantities of lensed GW+EM systems and supernovae. And next, let's look at the exact expressions for the time delay distance from the observational and theoretical perspectives, respectively.
\subsection{$D_{\Delta t}$ in theory} 
A homogeneous and isotropic universe can be described by the FLRW metric:
\begin{equation}\label{ds}  
ds^2 = - c^2 dt^2 + \frac{a(t)^2}{1-K r^2} dr^2 + a(t)^2 r^2 d\Omega^2. 
\end{equation}
 Note that the cosmic curvature $k=K/(H_0)^2=-\Omega_k$, where $\Omega_k$ represents the spatial curvature density parameter, and $H_0$ is the Hubble constant.  The dimensionless distance $d(z_l,z_s)=(1+z_s)H_0D_A(z_l,z_s)/c$ can be interpreted as:
\begin{equation} \label{dzz} 
d(z_l,z_s)=\frac{1}{\sqrt{-k}}{\rm sinh}(\sqrt{-k}\int_{\rm t_s(z_s)}^{\rm t_l(z_l)}\frac{H_0}{a(t)}\,dt).
\end{equation}
And according to the dimensionless distance, we have:
\begin{equation}\label{dlst}
\frac{d_{ls}}{d_ld_s}=\frac{c}{H_0(1+z_l)}\frac{D_A(z_l,z_s)}{D_A(z_l)D_A(z_s)}.
\end{equation}
We denote $d_{ls} = d(z_l,z_s), d_l = d(0, z_l)$  and  $d_s = d(0, z_s)$, where 0 means redshift to be 0. According to Eq.~\ref{dzz}, $d_{ls}$ can be re-defined in terms of $d_{l}$ and $d_{s}$ as following formula~\citep[namely, DSR, see][]{2015PhRvL...115...101301L}:
\begin{equation}  \label{dls}
d_{ls} = \epsilon _1d_s\sqrt{1-kd_l^2}-\epsilon _2d_l\sqrt{1-kd_s^2},
\end{equation}
where $\epsilon_i = \pm1$. For $k\leq0$,  $\epsilon_i = 1$. For $k\ge0$ the signs depend on the three-dimensional hypersphere location of the source, the lens and the propagation direction of the light~\citep{2014JCAP...03...035L}.
{Assuming t and z are in one-to-one relation and $d'(z)>0$, then $\epsilon_i = 1~$\citep{2015PhRvL...115...101301L}.
Later \citet{2017ApJ...839...70L} rewrote the DSR as follows:
\begin{equation}\label{dls2017}
\frac{d_{ls}}{d_ld_s} = T(z_l)-T(z_s),
\end{equation}
where 
\begin{equation} \label{tz}
T(z) = \frac{1}{d(z)}\sqrt{1-kd(z)^2}.
\end{equation}
In this paper, the comoving distances $d(z)$ are obtained from supernovae observations in a cosmological-model independent way (for details see Sec. \ref{sec:highlight}).
Therefore, by combining Eq.~(\ref{dt}),~(\ref{dlst}),~(\ref{dls2017}), the time delay distance can be theoretically expressed as: 
\begin{equation}\label{dt_th}
D_{\Delta t} = \frac{c}{H_0(1+z_l)(T(z_l)-T(z_s))}
\end{equation}

\begin{figure*}[tbp]
%\begin{minipage}[t]{0.5\textwidth}
%\centering
%\includegraphics[width=8cm]{final0051.eps}
%\end{minipage}	
%\begin{minipage}[t]{0.5\textwidth}
%\centering
\includegraphics[width=19cm]{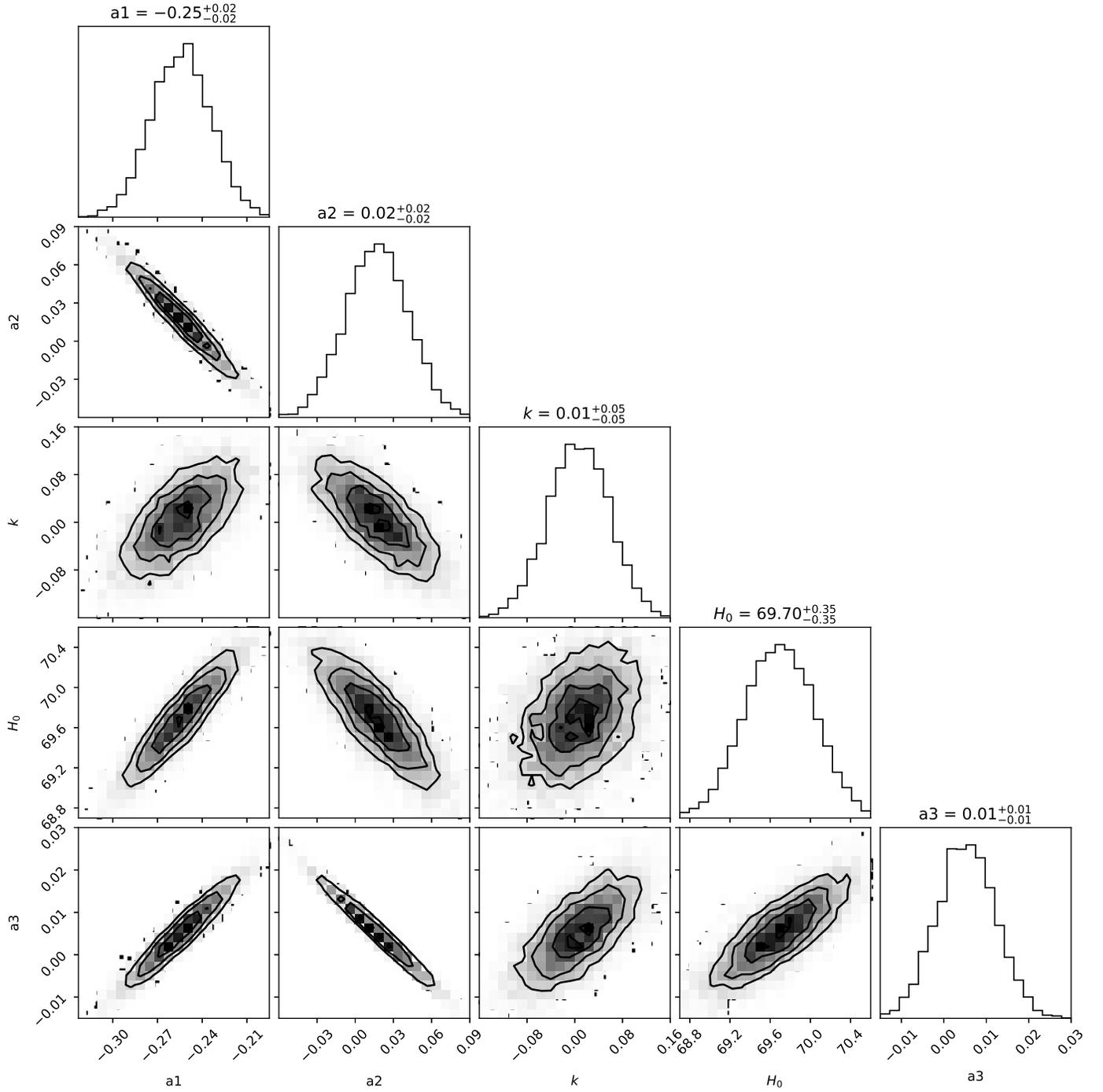}
%\end{minipage}
\caption{The one-dimensional marginalized distributions and the constraint contours with 5000 realizations for  $(a1, a2, a3, k, H_0)$ for 10 lensed GW+EM systems. The black solid line in each contour represents $1\sigma$, 2$\sigma$, 3$\sigma$ confidence interval, respectively. The numerical uncertainties at 1$\sigma$ confidence level are presented on the top of the figure.}
\end{figure*}

%####################
\subsection{$D_{\Delta t}$ by observation} 
The time delay distance~$D_{\Delta t}$ and the lensing system observations can be linked by the following formula:
\begin{equation}\label{delt}
\Delta t_{i,j}= \frac{(1+z_l)D_{\Delta t}}{c}\Delta \phi_{i,j},
\end{equation}
where $\Delta t_{i,j}$ is the time delay between images of the lensed source obtained from lensed GW, $z_l$ is the redshift of the lens, and $\Delta \phi_{i,j}$ is the Fermat potential difference between image positions which can be obtained by the EM counterpart of lensed GW:
\begin{equation}
\Delta\phi_{i,j}=[\frac{(\theta_i -\beta)^2}{2}-\psi(\theta_i)-\frac{(\theta_j -\beta)^2}{2}+\psi(\theta_j)],
\end{equation}
where $\theta_i,\theta_j$ represents the position of images of the lensed source, $\beta$ is the source position and $\psi$ is the two-dimensional lensing potential which is related to the mass distribution of the lens.

\section{Simulations and Results} \label{sec:highlight}
In order to constrain the cosmological parameters, we did simulations for 10 lensed GW+EM systems and 300 lensed quasar systems. The latter is adopted in accordance with \citet{2017ApJ...839...70L}.
For both systems, the theoretical values of cosmological parameters $(k,H_0)$ and d(z) can be derived simultaneously in a model-independent way from the observed quantities ($z_l, z_s, \Delta t,\Delta \phi$) of lensing systems and $D_L$ of supernovae observations based on the time delay distance in the following steps.\\

(1) The comoving distances d(z) in Eq.\ref{tz} are assessed in a cosmological-model independent way.
The simulated d(z) data are  obtained  according to the following formula in the flat $\Lambda CDM$ model:
\begin{equation}\label{dz}
D_L(z)=(1+z)d(z),
\end{equation}
where luminosity distance $D_L$ of Type Ia supernovae  are  from  simulated 3540 supernovae based on the 10-field hybrid strategy~\citep{2012ApJ...735...152L} of DES\footnote{The Dark Energy Survey (DES) is an international and collaborative effort which will carry out a deep optical and near-infrared survey of 5000 $\rm deg^2$ of the South Galactic Cap using a new 3 $\rm deg^2$ Charge Coupled Device (CCD) camera to be mounted on the Blanco 4m telescope at the Cerro Tololo Inter-American Observatory (CTIO)~\citep{2012ApJ...735...152L}. The 10-field hybrid strategy consists of two deep fields and the three shallow fields from the 5-field hybrid strategy, plus additional shallow fields clustered around the Chandra Deep Field-South field, and it offers an attractive balance among all important considerations~\citep{2012ApJ...735...152L,2017ApJ...839...70L}.}.   The redshifts of these supernovae are  $z<1.7$, which are consistent with the lensed GW+EM systems.  We direct the reader to \cite{2017ApJ...839...70L}  for detail of those  simulated Type Ia supernovae. 

 Then the theoretical  d(z)  used in Eq.\ref{tz}  and  Eq. \ref{chi}  are assumed  following  a fourth order polynomial:
 \begin{equation}\label{dz2}
 d(z)=z+a1 z^2 +a2 z^3+a3 z^4.
 \end{equation}
There is no much difference as long as d(z) is more flexible than a second order polynomial~\citep{2015PhRvL...115...101301L,2017ApJ...839...70L}.
  
(2) For 10 simulated lensed GW+EM systems, we used a typical lensing system which consists of a distant GW source accompanied by its EM counterpart and a foreground elliptical galaxy to obtain the observed quantities ($z_l, z_s, \Delta t_{i,j},\Delta\phi_{i,j} $).   The SIS model is adopted as the universal lens model. Specifically, ($z_l,z_s$) of 10 lensed GW events are generated based on the expected redshift probability distribution functions~(PDFs) of NS-NS systems in \citet{2015JCAP...12...006L}. The redshift probability distribution functions~(PDFs) is calculated after considering the intrinsic merger rates of the whole class of double compact objects located at different reshifts~\citep{2013ApJ...779...72L}, the designed sensitivity of the ET, and the probability that individual GW signals from inspiralling double compact objects could be lensed by an early-type galaxy~\citep{2015JCAP...12...006L}.
The source redshift is cut by 2, covering the supernovae redshift range.  As for another quantity, the Fermat potential, \citet{2017NatCo...8.1148L} has found that with lensed GW+EM signals, the lens modeling would yield the Fermat potential with 0.6\% uncertainty according to current techniques, which will directly give a few percent systematic error to the time delay distance. Therefore we take 0.6\% as the uncertainty of the Fermat potential difference $\Delta\phi_{i,j}$. 
 
(3) In addition to  above  redshifts and the Fermat potential  difference, we still need time delay $\Delta t$ to finally simulate the observational time delay distance. The time delay obtained from lensed GW is supposed to be very accurate with ignorable error~\citep{2017NatCo...8.1148L}. In this paper, we simply assume that the uncertainty of time delay obtained from lensed GW is 0~\citep{2018MNRAS...472...2906L}. 
\begin{table*}[htbp]
  \centering
  \begin{tabular}{ccccccc}
      \hline
      \hline
       &  No. & $a1$ & $a2$ & $a3$ & $k$ & $H_0$\\
      \hline
      lensed quasar & 300 & $-0.24_{-0.04}^{+0.04}$ & $0_{-0.06}^{+0.06}$ & $0.01_{-0.02}^{+0.03}$ & $0_{-0.06}^{+0.05}$ & $69.86_{-0.53}^{+0.53}$ \\
      \hline
      lensed GW+EM & 10 & $-0.25_{-0.02}^{+0.02}$ & $0.02_{-0.02}^{+0.02}$ & $0.01_{-0.01}^{+0.01}$ & $0.01_{-0.05}^{+0.05}$ & $69.7_{-0.35}^{+0.35}$  \\
      \hline
      \hline
  \end{tabular}
  \caption{The 1$\sigma$ confidence interval of five cosmological parameters calculated by different systems. From the top to the bottom, 300 lensed quasar systems, 10 lensed GW+EM systems, respectively. The difference between these two systems lies in that the time delay uncertainty is $3\%$ and 0, respectively.\\}
  \end{table*} 

(4) Finally, we generated the observational time delay distance in the flat $\Lambda CDM$ model with matter density $\Omega_M$ = 0.3, k=0 and Hubble constant {\bf$H_0 = 70 km s^{-1} Mpc^{-1}$}, and then took into account the observational  error propagated from the Fermat potential difference  Eq.~\ref{delt} as discussed in step (2). 

(5) To compare with traditional lensed quasar systems, following \citet{2017ApJ...839...70L}, we also simulated 300 lensed quasar systems from the OM10 catalogue \citep{2010MNRAS...405...2579} which provides mock observations of upcoming Large Synoptic Survey Telescope (LSST) based on realistic distributions of quasars and elliptical galaxies, and we have taken the parameter set ($z_l,z_s$) for the same lensed systems in this paper. As for the Fermat potential in lensed quasar systems, we considered that its uncertainty is the same as in time delay measurements. The uncertainty of the time delay measurement of lensed quasar systems in the EM domain is $\sim3$\%~, which combined with the Fermat potential would result in $\sim5$\%~ uncertainty in the time delay distance~\citep{2017ApJ...839...70L}.

(6) Given above  simulated  data,  $(k,H_0, d(z))$ and their relative uncertainties can be  derived in a model-independent way based on the theoretical time delay distance Eq.~\ref{dt_th} by minimising the following equation: 
\begin{equation}\label{chi}
\chi^2= \sum_{i=1}^{10}\frac{[D_{\Delta t(th)}-D_{\Delta t(ob)}]^2}{\delta D_{\Delta t(ob)}^2}+
\sum_{i=1}^{3540}\frac{[d(z)_{ob}-d(z)_{th}]^2}{\delta d(z)_{ob}^2},
\end{equation}
where ``$th$" and ``$ob$" represent theory and observation, respectively.
This process is achieved by the minimization function in python.
Eventually the five parameters $(a1, a2, a3, k, H_0)$ can be fitted simultaneously.

As a demonstration, the constraint contour with 5000 realizations for $(a1, a2, a3, k, H_0)$ for 10 lensed GW+EM systems are presented in Figure 1.   Clearly, we could recover the injection values of $(k, H_0)$ within a certain uncertainty ($\sim 1\%$ ).  For comparison, we showed the 1$\sigma$ numerical uncertainty of five cosmological parameters obtained from 300 lensed quasars in the EM domain and from 10 lensed GW+EM systems both in Table 1. It can be seen clearly from Table 1 that all the five parameters (especially the Hubble constant) can be constrained more precisely with lensed GW+EM systems. To be specific, the constraining power of only 10 lensed GW+EM systems is comparable and even stronger than 300 lensed quasar systems. For example, the Hubble constant from lensed GW+EM systems is $H_0=69.7_{-0.35}^{+0.35}$ against $H_0=69.86_{-0.53}^{+0.53}$ simply from lensed quasars. In addition, the spatial curvature in the local universe obtained from lensed GW+EM systems is $k = 0.01^{+0.05}_{-0.05}$ which is comparable to the one from the lensed quasar systems, $k = 0.00^{+0.05}_{-0.06}$. Therefore, we can draw the conclusion that lensed GW+EM systems will place much more stringent constraint on cosmological parameters than lensed quasar systems.

\section{discussion and summary}
With the coming era of gravitational waves, we are excitedly looking forward to gaining some new insights on the unsolved astrophysical problems by multi-messenger systems. Interestingly, some researchers have considered the lensing effects on gravitational wave signals for advanced
detectors~\citep{2014PhRvD...90...062003L, 2017NatCo...8.1148L,2017PhRvD..95f3512B,2017PRL...118...091101,2014JCAP...10...080L,2017MNRAS...465...4634L,2013JCAP...10...022L}. Moreover, in \citet{2017PhRvL.118i1102F} and \citet{2017NatCo...8.1148L}, lensed GW+EM systems have been discussed in detail to be used as a powerful tool to provide stringent constraints on cosmological parameters. In this context we apply more precise time delay obtained from lensed GW to cosmology research, specifically, constraining cosmological parameters in FLRW metric. Our results are presented in Figure 1. And the comparison of the results obtained from lensed quasars and lensed GW+EM systems are both shown in Table 1. The results clearly showed that only 10 lensed GW+EM systems could give comparable and even better constrained parameters than 300 lensed quasar systems. In particular, the $1\sigma$ uncertainty range of the cosmic curvature parameter is $-0.04<k<0.06$.
% * <liyufeng315@gmail.com> 2018-09-18T14:55:55.300Z:
%
% ^.

This paper is a preliminary attempt to apply lensed GW+EM multi-messenger systems to cosmology research. We have proved that accurate time delay measurements from lensed GW is a new tool for precision cosmology, which can be widely applied in the future, such as the mass density slope of elliptical galaxies and its evolution with redshift, and dark matter substructure in galaxy-scale halos~\citep{2017NatCo...8.1148L,2009ApJ...699...2L}. 
In a word, more constrained parameters obtained from lensed GW+EM systems will make contributions to the future study of cosmology and thus to our understanding of the evolution of the Universe.

\acknowledgments
The authors would like to thank the referees for their valuable comments which considerably improved the original text, and  acknowledge valuable input  from K. Liao and J. Wei.   Y.L. and L.G. are supported by  the National Program on Key Research and Development Project through grant No. 2016YFA0400804, and by the National Natural Science Foundation of China with grant No. 11333005, and by the Strategic Priority Research Program of the Chinese Academy of Sciences through grant No. XDB23040100.  X. F. is supported by the National Natural Science Foundation of China under Grants (No.11673008 and 11633001), the Strategic Priority Program of the Chinese Academy of Sciences (Grant No. XDB 23040100) and Newton International Fellowship Alumni Follow-on Funding.

\end{document}